\documentclass[english, a4paper, 12pt]{article}
\usepackage{graphicx,psfrag}
\usepackage[T1]{fontenc}
\usepackage[margin=1in]{geometry}
\usepackage{graphicx, color}
\usepackage{epsfig}

\begin{document}

\title{Trends \& Risk Premia:\\ Update and Additional Plots}
\author{Tung-Lam Dao, Daniel Hoehener, Yves Lemp\'eri\`ere, \\
Trung-Tu Nguyen, Philip Seager \& Jean-Philippe Bouchaud
\\Capital Fund Management, \\ 
23 rue de l'Universit\'e, 75007 Paris, France}
\date{\today}

\maketitle

\begin{abstract}
Recently, our group has published two papers that have received some attention in the finance community. One is about the profitability of trend following strategies over 200 years, the second is about the 
correlation between the profitability of ``Risk Premia'' and their skewness. In this short note, we present two additional plots that fully corroborate our findings on new data.
\end{abstract}

Recently, our group has published two papers that have received some attention in the finance community. One is about the profitability of trend following strategies over 200 years \cite{trends}, the second is about the 
correlation between the profitability of ``Risk Premia'' and their skewness \cite{skewness1,skewness2}. In this short note, we want to present two additional plots that fully corroborate our findings on new data. 

\begin{figure}[!htb]
\centering
\includegraphics[width=10cm]{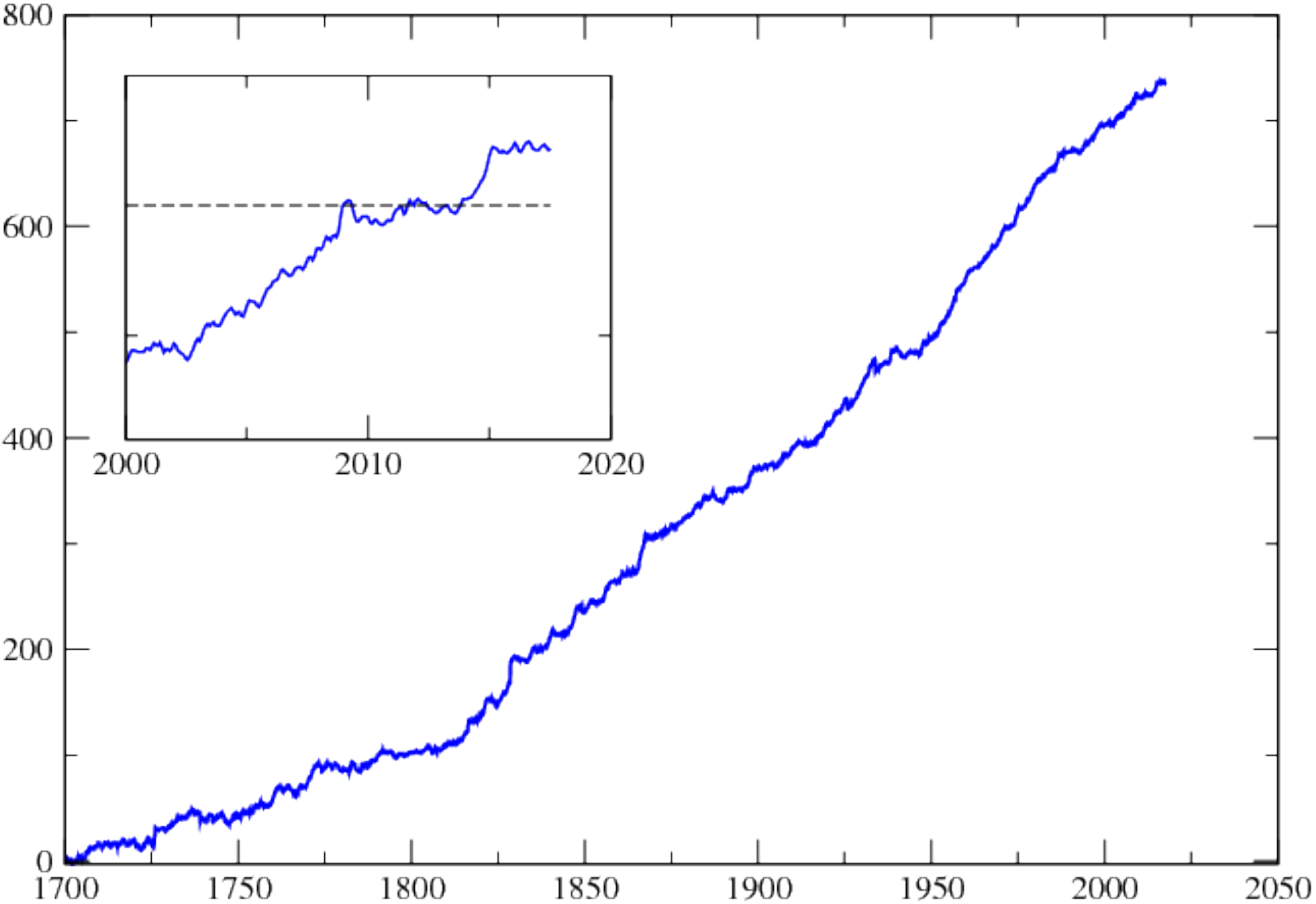}
\caption{Aggregate performance of a unit volatility portfolio exploiting a five months trend signal on all sectors since 1800. The period 1700 -- 1800 only contains the UK stock index, explaining the change of slope around 1800. Inset: Performance of trend-following in the recent period.}
\label{fig:trend}
\end{figure}

Fig.~\ref{fig:trend} shows the extension of the P\&L (Profit \& Loss) of the simple 5-month trend following strategy considered in \cite{trends}, both to a more recent period (2014 -- August 2017) and to an older period (1700 -- 1800), for which we managed to obtain 
the UK stock-index data. One clearly sees that trend-following on the UK stock index worked reasonably well in the 18th century, with a Sharpe ratio of around $0.3$, similar to the Sharpe ratio of trend following strategies on single indexes after 1800. The change of slope around 1800 comes from the fact that more and more contracts enter the pool after that date. The Sharpe ratio on our diversified pool of contracts since 1800 was found to be $\approx 0.7$ in \cite{trends}.

The second observation is that trend following strategies had been essentially flat in the five year period after the crisis (2009 -- 2014), and therefore declared ``dead'' in 2014 by many investors (precisely when our trend following paper first came out). Fig. \ref{fig:trend} not only shows that on a 300 year scale, this flat period is hardly observable, but also that the period 2015 -- 2017 has been perfectly in line with long term expectations. 

Turning now to the Risk Premia story, we recall that the main result of Refs. \cite{skewness1,skewness2} was that investors are not rewarded for taking volatility risk, but rather for taking skewness risk. The prime example is selling  options (i.e. insurance contracts). As is well known, this is a highly skewed strategy, since gains are limited while losses are unbounded. But it is also, historically, a very profitable strategy with a relatively high Sharpe, interrupted by strong dips when crises occur. Other so-called Risk-Premia strategies include the ``Carry Trade'' on currencies, fixed income portfolios, credit, and of course long the stock index (the so-called Equity Risk Premium). 
A scatter plot of Sharpe ratio $\rm{SR}$ vs. skewness $\zeta$ of a variety of different Risk Premia strategies revealed an approximately linear correlation between the two, given by:\footnote{We recall that the skewness measure used in  \cite{skewness1,skewness2} is a low-moment estimate, not the standard third-moment skewness.}
\begin{equation} \label{eq_line}
{\rm{SR}} = a - b \zeta, \qquad a \approx \frac13, \qquad b \approx \frac14,
\end{equation}
showing that the larger the (negative) skewness, the larger the Sharpe ratio. For example, shorting SPX options (at constant expected risk) generates a strategy with a Sharpe of $1.5$ and a skewness of $-4.6$. There are exceptions to this rule, for strategies that cannot reasonably be understood as ``Risk Premia'', such as trend-following itself that has both a positive Sharpe { and} a {\it positive} skewness (see \cite{convexity} for an extended discussion of this point).  

\begin{figure}[!htb]
\centering
\includegraphics[width=12cm]{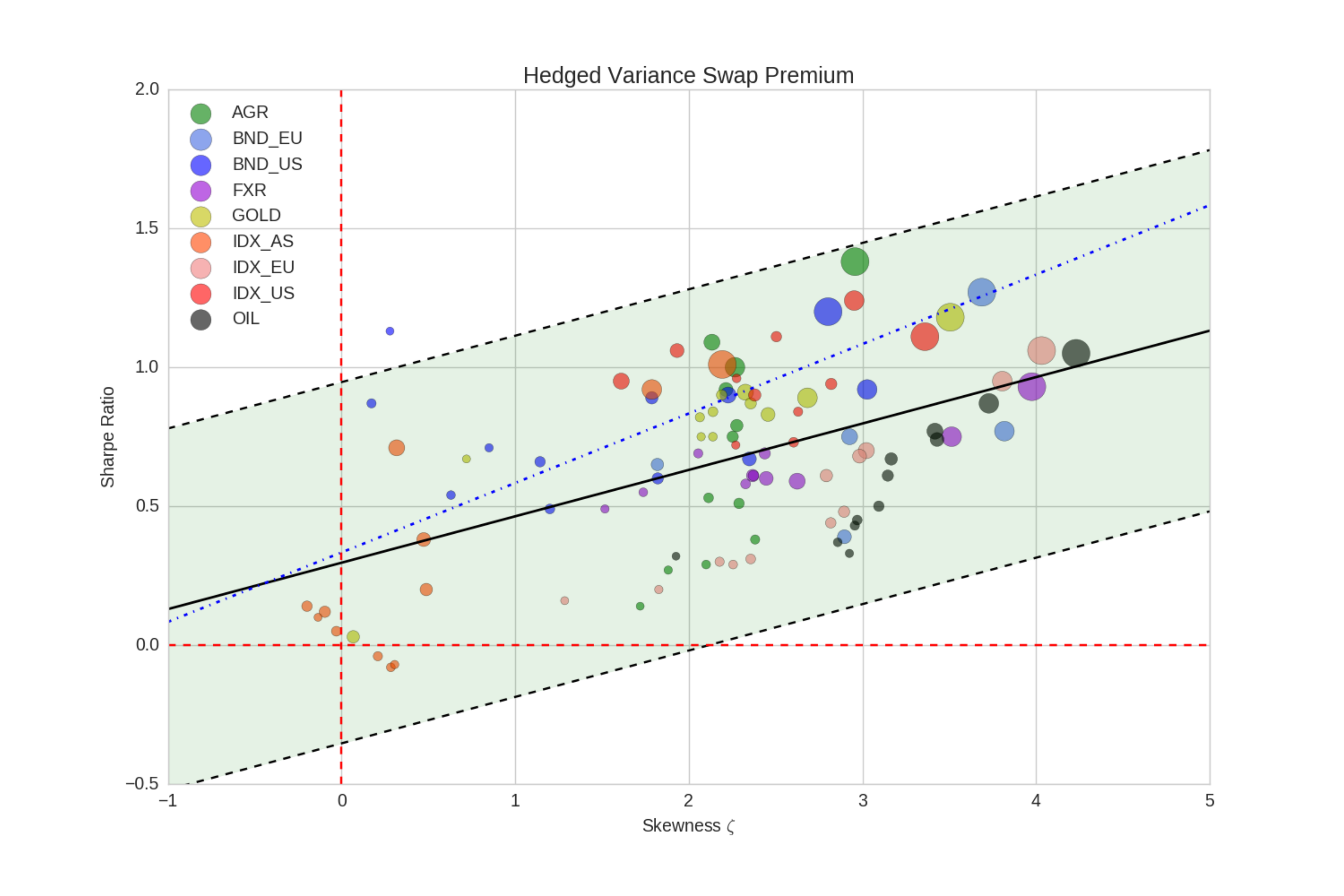}
\caption{Sharpe ratio versus (minus) skewness of hedged portfolios of uniform (across strikes) option strangles, for different maturities (1 to 12 months) and different underlyings, indicated in the legend. Circle sizes indicate the
maturity of the option, larger radii corresponding to shorter maturities. The plain line is a regression line through all the points, giving $a \approx 0.3$ and $b \approx 0.17$, close to the values found in Refs. \cite{skewness1,skewness2} (dash-dotted line).}
\label{fig:option_skew}
\end{figure}

We have now generated P\&L for shorting hedged variance swaps (in fact strike-uniform option strangles) with a fixed effective maturity between 1 month and 12 months, on a wide variety of underlying contracts. Compared to the data considered in \cite{skewness2}, where only 1 month maturity variance swaps were considered, we include options on US and EU bonds, and options on agricultural commodities. The resulting Sharpe vs. skewness plot is shown in Fig. \ref{fig:option_skew}, together with Eq. (\ref{eq_line}) for comparison. We see that all implied volatility premia seem to lie around the ``equilibrium'' line found in \cite{skewness1,skewness2}, vindicating the point of view that Risk Premia reflect some universal aversion to skewness. 

\begin{thebibliography}{10}

\bibitem{trends} Lemp\'eri\`ere, Y., Deremble, C., Seager, P., Potters, M., \& Bouchaud, J. P.  (2014). {\it Two Centuries of Trend Following}. Journal of Investment Strategies 3, 41-61.

\bibitem{skewness1} Lemp\'eri\`ere, Y., Deremble, C., Seager, P., Potters, M., \& Bouchaud, J. P.  (2015). {\it Tail risk premiums versus pure alpha}. Risk magazine.

\bibitem{skewness2} Lemp\'eri\`ere, Y., Deremble, C., Nguyen, T. T., Seager, P., Potters, M., \& Bouchaud, J. P. (2017). {\it Risk premia: Asymmetric tail risks and excess returns}. Quantitative Finance, 17(1), 1-14.

\bibitem{convexity} Dao, T.-L., Nguyen, T.-T., Deremble, C., Lemp\'eri\`ere, Y., Bouchaud, J.-P., Potters, M. (2016). {\it Tail protection for long investors: Trend convexity at work}, arXiv:1607.02410.

\end{thebibliography}
\end{document}